\documentclass[twocolumn,prl,aps]{revtex4}
\input{psfig.sty}
\input{epsf}    
\begin{document}
\title{Dynamical model of sequential spatial memory: 
winnerless competition of patterns}
\author{Philip Seliger, Lev S. Tsimring, Mikhail I. Rabinovich}
\affiliation{Institute for Nonlinear Science, University of California,
San Diego, La Jolla, CA 92093-0402}
\date{\today}
\begin{abstract}
We introduce a new biologically-motivated model of sequential 
spatial memory which is based
on the principle of winnerless competition (WLC). We implement this mechanism
in a two-layer neural network structure and present the learning
dynamics which leads to the formation of a WLC network. After learning,
the system is capable of associative retrieval of pre-recorded sequences
of spatial patterns.
\end{abstract}
\pacs{87.10.+e,05.45.-a,87.18.Sn,87.19.La}
\maketitle

%\section{Introduction}

It is well accepted that the hippocampus plays the central role in
acquisition and processing of information related to the representation
of physical space. The most spectacular manifestation of this role is
the existence of so called ``place cells'' which repeatedly fire when
an animal is in a certain spatial location
\cite{okeefe}. While much effort has been 
spent on experimental search and modeling of the so called ``cognitive map''
\cite{okeefe2} as a paradigm for spatial memory, 
recent neurophysiological research favors an alternative concept of 
spatial memory based on a linked collection of stored {\em episodes} 
\cite{eichenbaum}. Each  episode comprises a sequence
of {\em events} which, besides spatial locations, may include
other features of environment (orientation, odor, sound,
etc.). Each distinct event is accompanied by time-locked activity of a
certain hippocampal cell.  Dynamical modeling of the emerging concept of
the episodic memory is of apparent general interest for
neuroscience. Several models of associative sequential memory have
been proposed in the literature
\cite{kleinfeld}.
Most of them are based on the generalization of the Hopfield associative
memory network \cite{hopfield} to include asymmetric synaptic connections.
Accordingly, they suffer from difficulties typical for 
Hopfield-type networks: the abundance of spurious attractors
(sequences), complex structure of attractor basins, and sensitivity to
noise.

A dynamical model of the sequential spatial memory should be
based on the following experimental facts. First, there is a clear
separation between neurons directly responding to specific stimuli 
(we call them sensory neurons, SN) and hyppocampal cells in CA1 and 
CA3 regions (principal neurons, PN). PNs fire in response to a 
combined vector of stimuli 
corresponding to a particular event. Second, while 
sensory neurons are not directly connected to each other, the 
PNs are coupled via inhibitory connections controlled by interneurons
(INs).  Third, the synaptic connections among PNs and between PNs and
SNs exhibit Hebbian Long-Term Potentiation (LTP) \cite{bliss,markram}.  Based on
these features of the hippocampal network, we propose a two-layer
dynamical model of the  sequential spatial memory (SSM) that can answer
the following key questions.  (i) How is a certain event (e.g.
an image of environment) recorded in the structure of the synaptic connections
between multiple SNs and a single PN during learning?  (ii) What kind of the
cooperative dynamics forces individual PCs to fire sequentially, which
would correspond to a specific route (a sequence of scenes) in the
environment?  (iii) How complex should this network be to store a
certain number of different episodes without mixing different events or
storing spurious episodes?

The key mechanism of storing of sequential memories within our SSM model
is the {\em winnerless competition} (WLC) \cite{wlc}.  Every event
(image, odor, etc.) is represented by a fixed point in the phase space of
the model in which one PN is activated and others are silent.  Every
fixed point among those included in the sequence is a saddle with many
stable separatrices and a single unstable separatrix which connects it
with a fixed point representing the subsequent event.  The resulting
network of separatrices would form a complex heteroclinic connection
which would lead the system along the sequence of events in the
specific episode.  

Let us discuss the learning objectives which would lead to formation of
the sequential spatial memory (SSM).  The first objective is to learn
a projection map: as a result of unsupervised learning the image of a
particular environment (snapshot) encoded by heightened activity of the
group of sensory neurons (SNs) leads to the heightened activity (firing)
of just one PN (see Fig.1). The second objective is to learn the
temporal sequence of images. This can be achieved by modifying
inhibitory connections among PNs due to long-term potentiation (see e.g.
\cite{bliss}). The resulting structure of the phase space for the PN
layer will exhibit features of the winnerless competition \cite{wlc}.
After the learning is completed, the neural network should be able to
reproduce a specific route following a starting pattern. 

The two-layer structure of the SSM model is reminiscent of the
projection network implementation of the {\em
normal form projection algorithm} (NFPA) \cite{baird}. In that model, the dynamics
of the network is cast in terms of the normal form equations
which are written for amplitudes of certain normal forms which
correspond to different patterns stored in the system. The normal form
dynamics can be chosen to follow certain dynamical rules, for example,
in \cite{baird} it was shown that a Hopfield-type
network with improved capacity can be built using this approach.
Furthermore, in  \cite{baird} it was proposed that specific choices of
the coupling matrix for the normal form dynamics can lead to
multi-stability among more complex attracting sets than simple fixed
points, such as limit cycles or even chaotic attractors.  As we will see
below, the model of SSM after learning is completed can be viewed as a
variant of the NFPA with m specific choice of normal form dynamics
corresponding to the winnerless competition among different patterns. 

%\section{Model}
Consider a two-level network 
of $N_s$ sensory neurons (SN) $x_i$ and $N_p$ principal
neurons $a_i$. 
Similar to the 
projection network model \cite{baird}, we assume that sensory neurons do
not have their own dynamics and are slaved to either external stimuli
in the learning (or storing) regime, or to the PNs in the
retrieval regime. In the learning regime, $x_i=I_i$ where $\{I_i\}$ is a
binary input pattern consisting of 0's and 1's. During the retrieval
phase, $x_i=\sum_{j=1}^{N_p} P_{ij}a_j$,
where $P_{ij}$ is the $N_s\times N_p$ projection matrix of connections
between SNs and PNs.

The PNs are driven by SNs during the learning
phase, but they also have their own dynamics controlled by inhibitory
inter-connections (see above). After the learning is finished, the direct driving
from SNs is disconnected.
The equations for the amplitudes of PNs, $a_i$, read
\begin{equation}
\dot{a_i}= a_i-a_i\sum_{j=1}^{N_p}V_{ij}a_j+\alpha a_i\sum_{j=1}^{N_s}P_{ij}^Tx_j+\xi(t),
\label{a}
\end{equation}
where $\alpha \ne 0$ in the learning phase, and $\alpha=0$ in the retrieval
phase. We use the transposed projection matrix $P_{ij}^T$ assuming that the 
coupling between SNs and PNs is bi-directional and symmetric.
The last term in the r.h.s. of Eq.(\ref{a}) represents small positive external
perturbations which we model as white noise uniformly distributed between 0
and $\sigma$, however in reality it can represent input signals from other 
parts of the brain which control learning and retrieval dynamics. 

After a certain pattern is presented to the model, the sensory
stimuli reset the state of the PN layer according to the projection rule
$a_i=\sum_{j=1}^{N_s} P_{ij}^Tx_j,$
but then $a_i$ change according to Eq.(\ref{a}).

In addition to the dynamics of SNs and PNs during learning and retrieval
phases, we need to introduce two learning processes: (i) forming the
projection matrix $P_{ij}$ which is responsible for connecting a group
of sensory neurons of the first layer corresponding to a certain stored
pattern to a single principal neuron which represents this pattern at the
PN level; (ii) learning of the competition matrix $V_{ij}$ which is
responsible for the temporal/logical ordering of the sequential memory.

{\em Projection matrix.} The slow learning dynamics of the projection matrix is
controlled by the following equation
\begin{equation}
\dot P_{ij}=\epsilon a_i(\beta x_j-P_{ij}).
\label{Pij}
\end{equation}
with $\epsilon\ll 1$.
We assume that initially all $P_{ij}$ connections are nearly identical 
$P_{ij}=1+\eta_{ij}$, where
$\eta_{ij}$ are small random perturbations, $\sum_j \eta_{ij}=0,\
\langle\eta_{ij}^2\rangle=\eta_0^2\ll 1$. Additionally, we assume that
initially matrix $V_{ij}$ is purely competitive: $V_{ii}=1$ and
$V_{ij}=V_0>1$ for $i\ne j$. 

\begin{figure}[h]
\centerline{ \psfig{figure=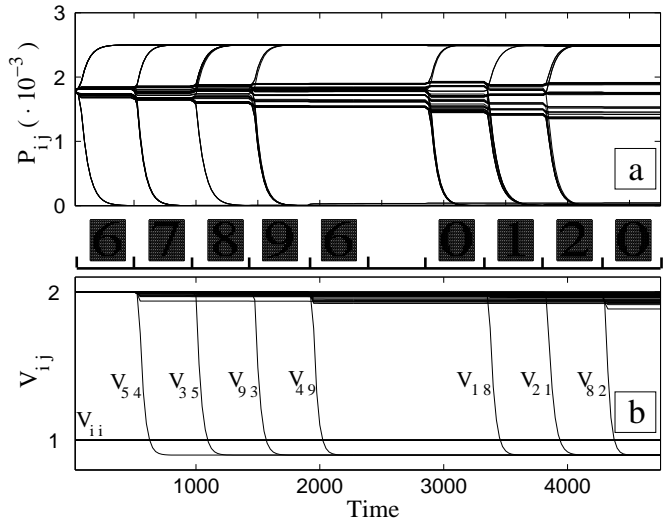,width=3.5in}}
\caption{The strengths of the connection coefficients between the
sensory
and the principal layers (a) and within the principal layer (b).
Parameters of simulations: $N_s=588, N_p=10, \alpha=1, \beta=2.5, V_1=0.9, 
\epsilon=0.01, \sigma=10^{-4}, \tau=480$
}
\label{fig1}
\end{figure}

Consider a scenario when we want to ``memorize'' a certain pattern {\bf
A} in our projection matrix. We apply a set of inputs $A_i$
corresponding to the pattern {\bf A} to the SNs. As before,
we assume that external stimuli render the SNs in one of two
states: excited ($A_i=1$) and quiescent ($A_i=0$). The initial state of
the PN layer is fully excited ($a_i(0)=\sum_j P_{ij}A_j$).
According to the competitive nature of interaction of PNs
after a short transient, only one of them (neuron A) which
corresponds to maximum $a_i(0)$ remains excited and others become
quiescent (inhibited). Which neuron becomes ``responsible'' for the
pattern {\bf A} is actually random, as it depends on the initial
projection matrix $P_{ij}$. As it follows from Eq.(\ref{Pij}), for small
$\epsilon$ `synapses' of suppressed PNs don't change,
whereas synapses of the (single) excited neuron evolve such that the
connections between excited SNs and PNs neurons amplify
towards $\beta>1$, and connection between excited PNs and quiescent
SNs decay to zero (see Figure \ref{fig1},a). As a result,
the first input pattern will be ``recorded'' in one of the rows of the
matrix $P_{ij}$, while other rows will remain almost unchanged.

Now suppose that we want to record a second pattern different from the
first one. We can repeat the procedure described in the previous
paragraph, namely, apply external stimuli (pattern {\bf B}) to the
SNs, ``project" them to the initial state of the PN
layer ($a_i(0)=\sum_j P_{ij}B_j$), and let the system evolve.  Since
synaptic connections from SNs suppressed by the first
pattern to neuron A have been eliminated, a new set of stimuli
corresponding to pattern {\bf B} will excite neuron A weaker than most
of the others, and competition will lead to selection of one principal
neuron B {\em different} from neuron A.  In such a way we can record as
many patterns as there are PNs.

{\em Competition matrix}. The sequential order of patterns recorded in the
projection network is determined by the competition matrix
$V_{ij}$. Initially it is set to $V_{ij}=V_0>1$ for $i\ne j$ and
$V_{ii}=1$ which corresponds to winner-take-all competition. The goal of
sequential spatial learning is to record the transition of pattern {\bf A} to pattern
{\bf B} in the form of suppressing the competition matrix element $V_{BA}$.
The slow dynamics of the non-diagonal elements of the competition
matrix are controlled by the delay-differential equation
\begin{equation}
\dot V_{ij}=\epsilon a_i(t)a_j(t-\tau)(V_1 -V_{ij}).
\label{Vij}
\end{equation}
As seen from Eq.(\ref{Vij}), only the matrix elements corresponding to
$a_i(t)\ne0$ and $a_j(t-\tau)\ne0$, are changing towards the
asymptotic value $V_1<1$ corresponding to the desired transition.  Since
most of the time (except for short transients) only one of the principal
neurons is excited, only one of the connections $V_{ij}$ is changing at
any time (see Figure \ref{fig1},b). As a result, an arbitrary (non-repeating) sequence of patterns
can be recorded. If, after a series of non-repeating patterns, we show
the first pattern again, the ``loop'' of heteroclinic connections will
be closed and the system will be able to reproduce a repeating sequence
of patterns in a cyclic manner. 

If the dimension of the secondary layer $N_s$ permits, it is easy to
record into the network more than one sequence of patterns. To avoid a
spurious connection between the sequences, the time interval between the
last pattern of the first sequence and the first pattern of the second
sequence should be greater than $\tau$. 

In Figure \ref{fig1} we
show the simulation results for a slow dynamics of weights $P_{ij}$ and 
$V_{ij}$ during a
learning phase in a network with 588 sensory and 10 principal neurons
for $\epsilon=0.01$.  
As stored patterns $I_i$ we take ten digits 0,...9 represented as 21x28 pixel
dithered images. 
Two loop sequences of patterns have been
presented: ``0'', ``1'', ``2'', and ``6'', ``7'', ``8'', ``9''. Note that these
images are not precisely orthogonal to each other, and yet the system is able to
associate them to different PNs. While a certain
pattern is presented to the SN layer, certain matrix coefficients
$P_{ij}$ decay, some other (connecting excited neurons of the sensory
layer and a single excited neuron of the PN layer) approach 2.5
and remaining connections remain almost unchanged. After a switch from
one pattern to the next in a sequence the corresponding matrix coefficient
$V_{ij}$ decays to a low value $V_1=0.9$. 

\begin{figure}[h]
\centerline{ \psfig{figure=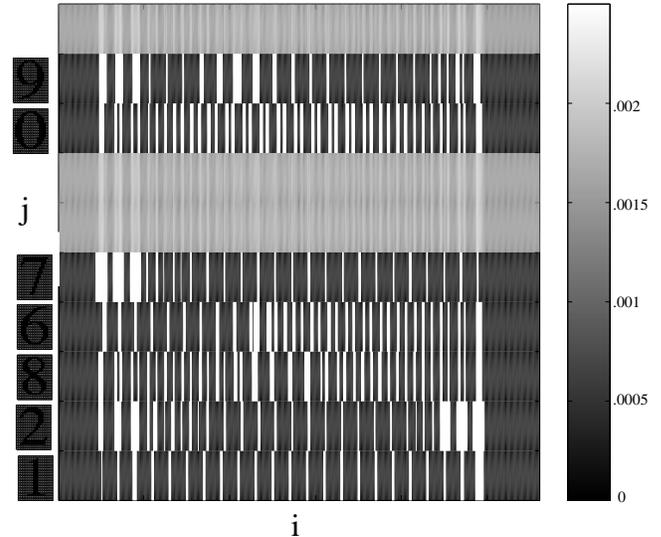,width=3.5in}}
\caption{The gray-level coded structure of the projection matrix $P_{ij}$ 
after seven distinct patterns have been recorded. Black-and-white rows
correspond to recorded patterns (shown on the right) and gray rows
corresponds to available PNs. Parameters of simulations are the same as
in Fig.\protect\ref{fig1}.
}
\label{fig2}
\end{figure}

Figure \ref{fig2} shows the state of the projection matrix after all
seven
patterns have been recorded into the memory. The high-contrast ``barcode'' rows
correspond to memorized patterns and dim gray rows correspond to yet
available principal neurons.

\begin{figure}[h]
\centerline{ \psfig{figure=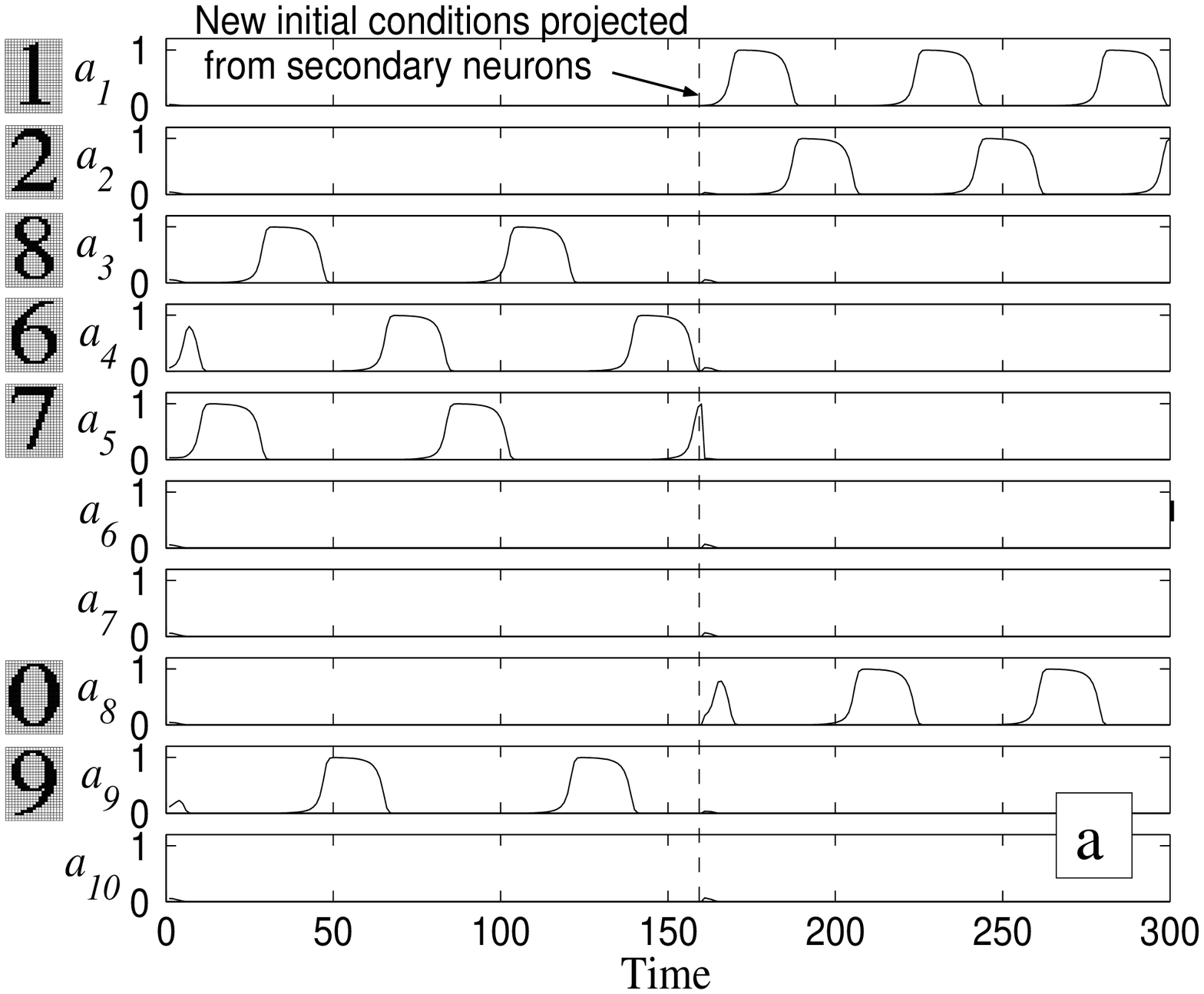,width=3.in}}
\centerline{ \psfig{figure=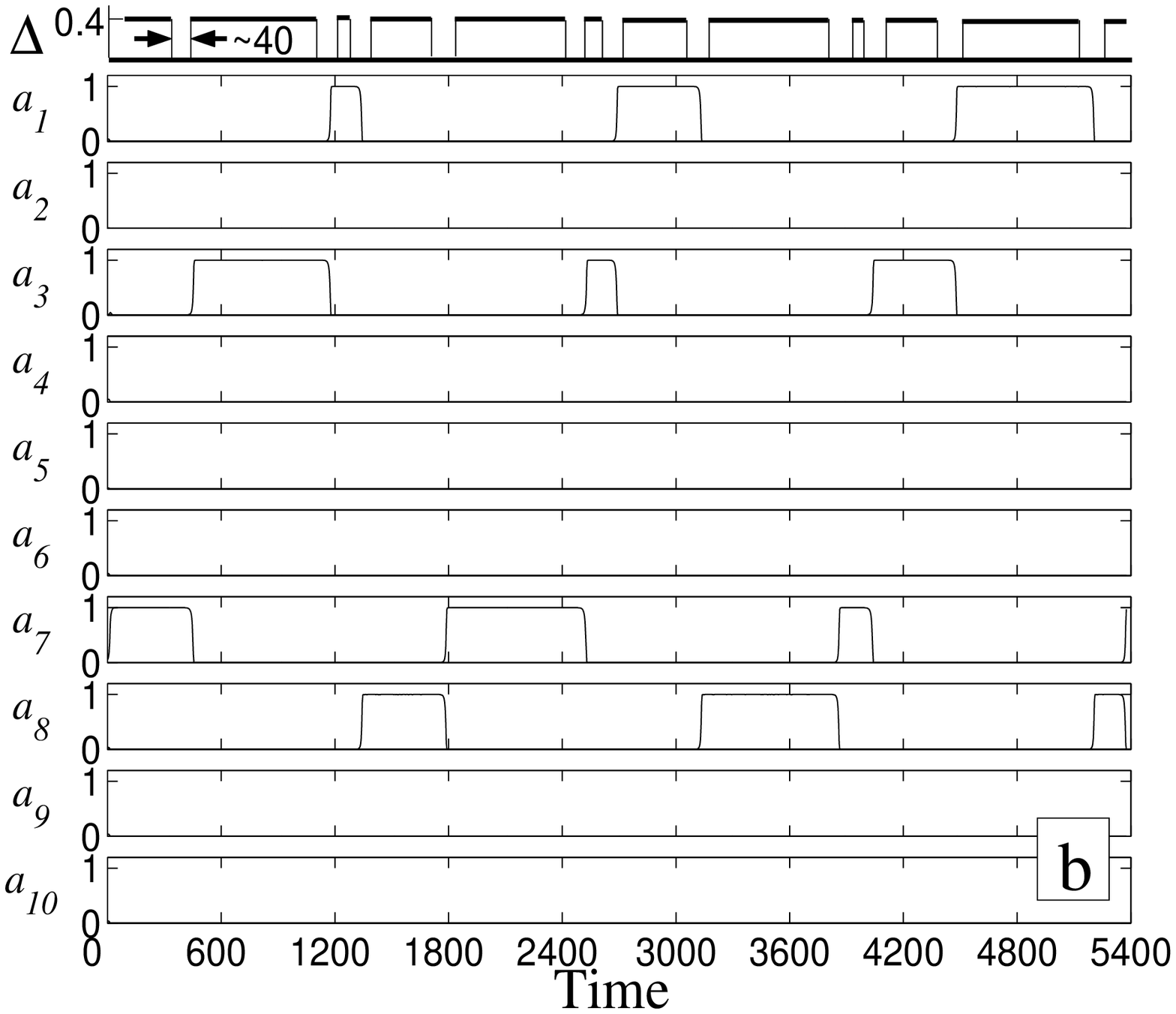,width=3.in}}
\caption{
Amplitudes of principal neurons during the memory retrieval phase, 
a - periodic retrieval, two different test patterns presented,
b - aperiodic retrieval with modulated inhibition.
Parameters of simulations are the same as
in Fig.\protect\ref{fig1}.
}
\label{fig3}
\end{figure}

Now, presenting a test pattern {\bf T} ``resembling'' one of the
recorded patterns to the sensory layer ($x_i(0)=T(i)$, $a_i(0)=\sum_i
P_ij^T T_j$), will initiate a periodic sequence of patterns
corresponding to the previously recording sequence recorded in
the network. 
Figure \ref{fig3},a shows the behavior of the principal neurons after two
different initial patterns have been presented, one resembling digit
``0'' and another resembling digit ``6''. In both cases, the system
quickly settled onto a cyclic generation of patterns associated with a
given test pattern.  At any given time except for a short
transient time between the patterns, only a single principal neuron is
``on'', which corresponds to a particular pattern. The order in which the
principal neurons  are turned on is completely determined by the
structure of the WLC matrix $V_{ij}$. The duration of
each ``state'' is determined by the magnitude of external perturbations
$\sigma$. For $\sigma=0$, the system would asymptotically approach the
separatrices and so the durations of each state would grow indefinitely.
For a finite $\sigma$, the duration of each state scales as
$-\log\sigma$. 
In the above example, patterns are retrieved from the spatial memory
periodically in time.  However, it may be desirable for a system to be
able to control the duration of individual patterns in the sequence.
This can be easily achieved by modulating  the overall strength of
inhibitory connections $V_{ij}+\Delta(t)$. While $\Delta(t)>1-V_1$, all
fixed points are stable nodes, and so a single principle neuron keeps
firing. In order to advance to the next patter, $\Delta(t)$ is
suppressed to zero for a short period of time ($O(-\log\sigma)$). An
example of such non-periodic retrieval of images is shown in Fig.
\ref{fig3},b.

In conclusion, we demonstrated the new principle of operation for the
sequential spatial memory which is based on the winnerless competition.
This principle is embodied in the two-layer neuronal structure with the
first layer serving as a sensory input for the second layer which
performs winnerless competition among representative principal neurons.
We introduced the learning rules for the projection and the competition
matrices which lead naturally to the desired function of the network. We
also demonstrated that external perturbations can influence the
timing of the transitions among the stored patterns, however, the
sequence of patterns is robust against external perturbations. The model
can operate in two regimes: externally-timed  switching controlled by global
modulation of inhibitory connections and spontaneous periodic switching
between patterns.  The latter can be relevant for route replays during
sleep.  Of course, our model only describes a generic mechanism of
sequential spatial memory, in real biological systems neurons generate
non-stationary spike trains and synaptic dynamics is timing-dependent
(see \cite{markram,poo}). Moreover, instead of a single PN a given
pattern can be represented by a group of neurons which would increase
the structural stability of the memory. All these generalizations will
be addressed in our future work. 

Authors gratefully acknowledge support from the Engineering Research
Program of the Office of Basic Energy Sciences at the U.S. Department of
Energy, grant DE-FG03-96ER14592 and from NSF grant EIA-013708.

\end{document}